\documentclass{astlb}

\usepackage[utf8]{inputenc}
\usepackage[english]{babel}

\usepackage{caption}
\usepackage{color}

\usepackage{amsmath}
\usepackage{amssymb}

\usepackage{graphicx}
\usepackage{multirow}
\usepackage{rotating}
\usepackage{lscape}
\usepackage{caption}
\usepackage{subcaption}
\usepackage{nameref}
\usepackage{fancyref}
\usepackage{mathtools}
\usepackage{gensymb}
\usepackage{array}
\usepackage{tabularx}
\usepackage{adjustbox}
\usepackage[hyperfootnotes=false]{hyperref}

\newcolumntype{C}{>{\small\centering\arraybackslash}X}

\newcommand {\be}{\begin {equation}}
\newcommand {\ee}{\end {equation}}

\def\ergs{erg~s$^{-1}$}
\def\*{$^{*}$}

\sloppypar

\begin{document}

\journalinfo{2018}{44}{3}{1}[13]

\title{PECULIARITIES OF SUPER-EDDINGTON FLARES \\ FROM THE X-RAY PULSAR LMC X-4 WITH \slshape{NuSTAR}}

\author{\bf A.E.~Shtykovsky\email{a.shtykovsky@iki.rssi.ru}\address{1}, V.A.~Arefiev\address{1}, A.A.~Lutovinov\address{1}, S.V.~Molkov\address{1}
\addresstext{1}{Space Research Institute, Moscow, Russia}}

\shortauthor{Shtykovsky et al.}

\shorttitle{PECULIARITIES OF SUPER-EDDINGTON FLARES FROM LMC\,X-4}
\submitted{27.06.2017}

\begin{abstract}
\noindent
We present results of the analysis of super-Eddington flares registered from the X-ray pulsar LMC\,X-4 by the \textsl{NuSTAR} observatory in the broad energy range 3--79~keV. The pulsar spectrum is well described by the thermal comptonization model ({\sc comptt}) both in a quiescent state and during flares, when the peak luminosity reaches values $L_{\rm x} \sim (2-4)\times10^{39}$\,\ergs. An important feature, found for the first time, is that the order of magnitude increase in the luminosity during flares is observed primarily at energies below 25--30~keV, whereas at higher energies (30--70~keV) the shape of the spectrum and the source flux remain practically unchanged. The increase of the luminosity is accompanied by changes in the source pulse profile -- in the energy range of 3--40~keV it becomes approximately triangular, and the pulsed fraction increases with increasing energy, reaching 60--70\,$\%$ in the energy range of 25--40~keV. The paper discusses possible changes in the geometry of the accretion column, which can explain variations in spectra and pulse profiles.


\end{abstract}

\section*{Introduction}

\begin{table*}[t]
\vspace{6mm}
\centering
\caption{\textsl{NuSTAR} observations of LMC~X-4.} \label{tab:obsphase}
\begin{tabular}{c|c|c|c|c} \hline
ObsID & $\Psi_{\rm orb}$ & $\Psi_{\rm sup}$ & Start time (MJD) & Exposition, sec \\
\hline
30102041002 & [0.29; 0.63] & [0.88; 0.89] & 2015-10-30 01:01:08 (57325.0425) & 24551 \\
30102041004 & [0.39; 0.69] & [0.07; 0.08] & 2015-11-04 19:46:08 (57330.8237) & 21880 \\
30102041006 & [0.11; 0.43] & [0.28; 0.30] & 2015-11-11 11:16:08 (57337.4695) & 22986 \\
30102041008 & [0.42; 0.73] & [0.81; 0.82] & 2015-11-27 09:16:08 (57353.3862) & 20282 \\
\hline 

\end{tabular}
\end{table*}

\noindent
LMC X-4 is a high-mass X-ray binary system located in the Large Magellanic Cloud (an estimated distance is $d = 50\text{ kpc}$), with a spin period of $P_ {\rm spin} \sim 13.5 \text{ sec}$ and an orbital period of $P_{\rm orb} \simeq 1.4 \text{ days}$. The system consists of a neutron star with the mass of $M_{\star} \simeq 1.57M_{\sun}$, where $M_{\sun}$ is the solar mass, and an optical companion, the O8III star with the mass of $\sim 18 M_{\sun}$ \citep[see][and references therein]{2015A&A...577A.130F}. X-ray observations show the presence of X-ray eclipses in the LMC\,X-4 system \citep{1978Natur.271...37L,1978Natur.271...38W} and variations of an X-ray flux with a period of $P_{\rm sup} \simeq 30.5 \text{ days}$ (the so-called superorbital period), commonly associated with the precessing accretion disk, which periodically obscures the direct emission from the neutron star \citep{1981ApJ...246L..21L,1989A&A...223..154H,2009ApJ...696..182N}. The observed X-ray flux change during the super-orbital cycle by a factor of $\sim 50$ \citep{1989A&A...223..154H}. The intrinsic luminosity of the pulsar remains approximately constant (taking into account local fluctuations) with the value of $L_{\rm x} \sim (2 - 5) \times 10^{38}$\,\ergs \citep[see][and references therein]{2000ApJ...541..194L,2005AstL...31..380T,2015AstL...41..562M,2017AstL...43..175S}, which corresponds or slightly exceeds the Eddington limit for a neutron star with the mass of $\sim 1.57M_{\sun}$.

In close X-ray binaries with compact objects it is not uncommon to observe transient increases in X-ray luminosity, which are called flaring events. These events may be of different nature -- in particular, flares may occur due to thermonuclear flashes on the surface of the neutron star, non-stationary accretion, etc.

Flaring activity of LMC\,X-4 has been observed in X-rays since its discovery \citep{1983ApJ...264..568K}. It should be noted that flares has been registered both in high and low states \citep{1996ApJ...467..811W}. Flares are observed as an episodical super-Eddington luminosity events ($\sim 10^{39}-10^{40}$\,\ergs) lasting up to few thousand seconds \citep[see e.g.][]{2000ApJ...541..194L,2003ApJ...586.1280M}. This activity doesn't bear regular occurrences and, apparently, has an aperiodic character \citep{2000ApJ...541..194L}. Light-curves during the flares are modulated with the spin period \citep{1983ApJ...264..568K,2000ApJ...541..194L,2001ApJ...549L.225M}. Thus far there hasn't been revealed any plausible exhaustive physical theory (or mechanism) that explains the nature, energy release and timing properties of LMC\,X-4 flares. The study of the flaring activity is of utmost interest in the context of the instrinsic physical processes of the system. It involves the accumulation of significant amounts of energy and nonstationary processes evolution, leading to release of that energy in short time. The associated dynamical activity of the pulsar is the subject of interest as well. Especially intriguing aspect is the very high luminosity of the source during the flares (up to $\sim 10^{40}$\,\ergs), which may allow one to attribute LMC\,X-4 to a class of a flaring ultraluminous sources (FULX).

In the present paper a detailed analysis of the pulsar activity in various orbital and superorbital phases has been done for the first time using the \textsl{NuSTAR} observatory data. Special attention has been paid to the X-ray flaring activity detected in some of the observations.

\section*{Observations and data reduction}

\noindent
This paper analyses data, obtained by the \textsl{NuSTAR} \citep{2013ApJ...770..103H} during the observations of LMC\,X-4 in October and November of 2015 (ObsIDs 30102041002, 30102041004, 30102041006 and 30102041008). These observations cover various phases of the orbital ($\Psi_{\rm orb}$) and superorbital ($\Psi_{\rm sup}$) cycles of the system \citep[according to ephemeris from][]{2015AstL...41..562M}. An exposure time in science mode (Mode~01) in each observation was about $20-25 \text{ ksec}$. The log of observations is summed in Table~\ref{tab:obsphase}.

The primary data reduction has been carried out by the standard \textsl{NuSTAR} data analysis software ({\sc nustardas}, version 1.5.1) with the {\sc caldb} calibration database (version 20160922). The subsequent data reduction and analysis were performed with the {\sc heasoft} software package (version 6.19). The photon arrival times were corrected to the Solar system barycenter using the standard {\sc nustardas} tools. The corresponding correction of the photon arrival times for the motion of the neutron star in the binary system was made using the orbital parameters from \citet{2015AstL...41..562M}.

To obtain the source lightcurve, raw lightcurves from each of \textsl{NuSTAR} modules (FPMA and FPMB) has been background subtracted and an orbital correction has been applied \citep[following][]{2015ApJ...809..140K}. The resulting lightcurves were combined using the {\sc lcmath} utility of the {\sc heasoft} package.

We searched for the pulse period using the epoch-folding technique (the {\sc efsearch} utility of the {\sc heasoft} package). Pulse profiles were constructed by folding the source’s light curve with the period found. The energy spectra were analyzed using the {\sc xspec} package (version 12.8).

\bigskip

\section*{Results}

\subsection*{Lightcurves}

\begin{figure*}
\centering
\includegraphics{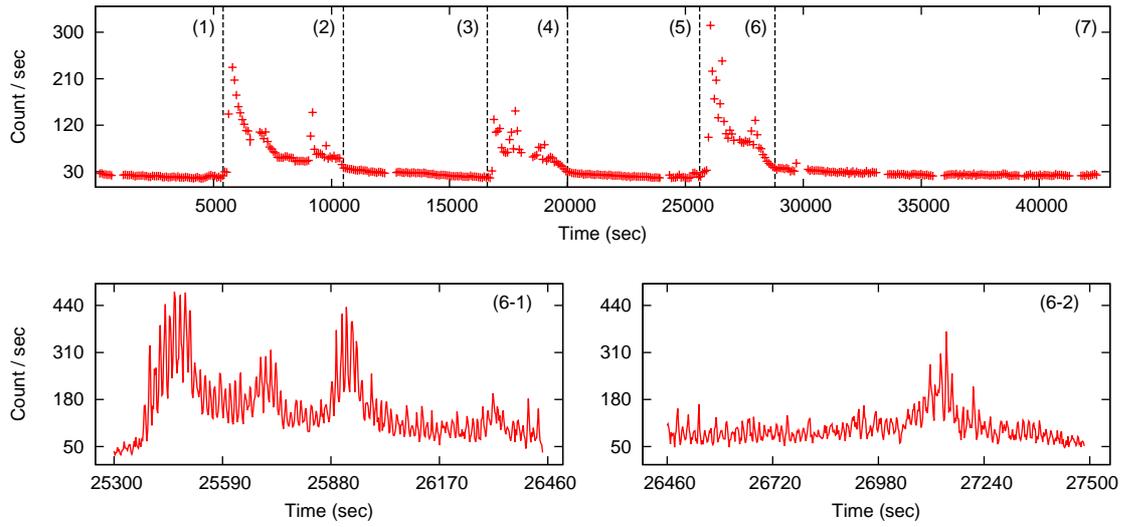}
\caption{(top panel): lightcurve of LMC\,X-4 in the observation 30102041002; (bottom panels): lightcurve for the FL3 event in the observation 30102041002 (interval 6): main flare (bottom left panel, 6-1) and afterglow (bottom right panel, 6-2). See the text for the detailed description. Zero time corresponds to the beginning of observation (Table \ref{tab:obsphase}).}
\label{fig:lc02}
\end{figure*}

\begin{figure*}
\centering
\includegraphics{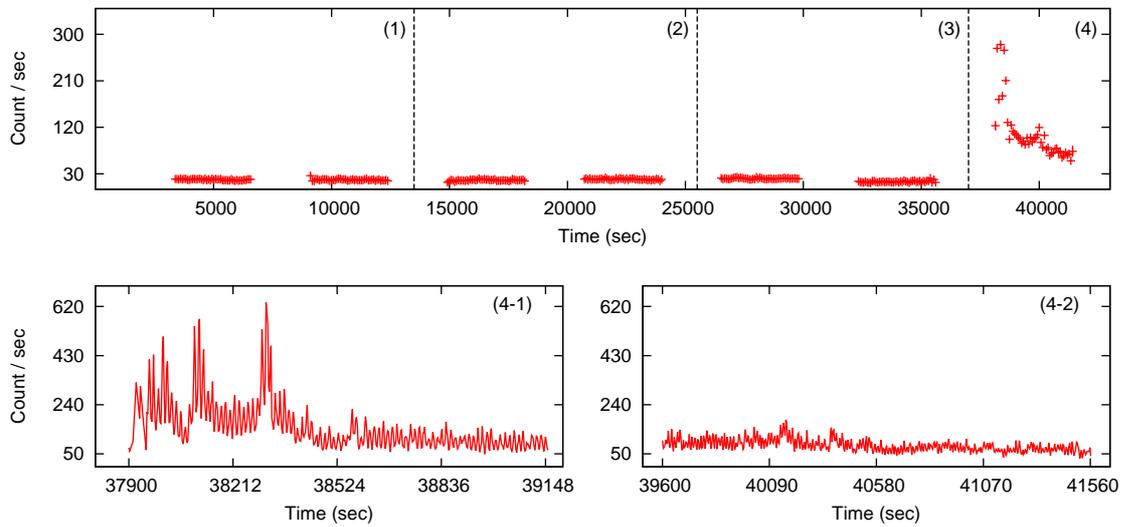}
\caption{(top panel): lightcurve of LMC\,X-4 in the observation 30102041008; (bottom panels): lightcurve for the FL4 event in the observation 30102041008 (interval 4): main flare (bottom left panel, 4-1) and afterglow (bottom right panel, 4-2). See the text for the detailed description. Zero time corresponds to the beginning of observation (Table \ref{tab:obsphase}).}
\label{fig:lc08}
\end{figure*}

\noindent
Pulsar lightcurves in the energy range 3--79~keV for observations 30102041002 and 30102041008 are presented in Fig.~\ref{fig:lc02} and Fig.~\ref{fig:lc08}, correspondingly. There are two episodes of flaring activity in these observations: one contains three flaring events (FL1, FL2 and FL3); another -- one flaring event (FL4). Events FL1~--~FL3 are separated by $\sim 10 \text{ ks}$ from each other. Episodes of activities in observations 30102041002 and 30102041008 (events FL1 and FL4) are within the $\sim 28.3 \text{ days}$. The duration of each event is about of $4000 \text{ sec}$. All events have a similar morphology: fast rise, exponential decay (characteristic time $t_{\rm exp} \sim 800-1200 \text{ sec}$), afterglow, strong pronounced pulsations on the pulsar spin frequency and presence of subpulses during the main flare and the afterglow (see lower panels in Fig.~\ref{fig:lc02} and Fig.~\ref{fig:lc08}). In observations 30102041004 and 30102041006 no flaring activities has been detected.

For subsequent analysis all observational data, containing flares, were split into several intervals. The intervals were chosen to separate flaring and quiescent states (see upper panels in Fig.~\ref{fig:lc02} and Fig.~\ref{fig:lc08}). Each flaring event was in turn split into a main flare and an afterglow parts (see lower panels in Fig.~\ref{fig:lc02} and Fig.~\ref{fig:lc08}). Observational data for 30102041004 and 30102041006 were analyzed as a whole.

\bigskip

\subsection*{Spin period and pulse profiles}

\begin{figure*}
\centering
\begin{subfigure}{0.4\textwidth}
\includegraphics{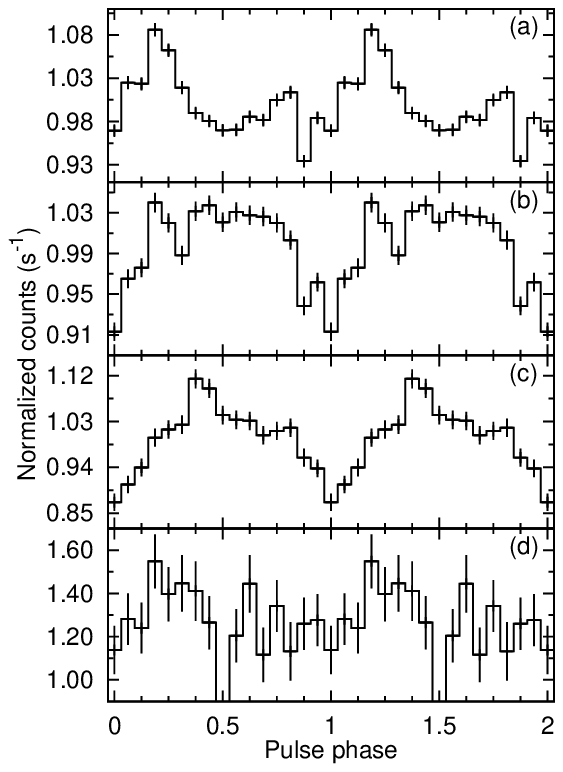}
\end{subfigure}
\begin{subfigure}{0.4\textwidth}
\includegraphics{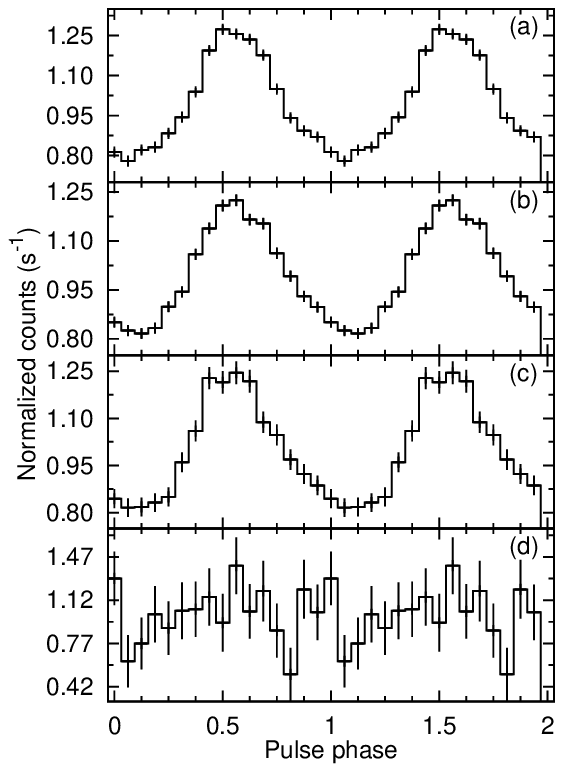}
\end{subfigure}

\caption{Pulse profiles of LMC\,X-4 in the quiescent state: observations 30102041004 (left panel) and 30102041006 (right panel) in the energy ranges: 3--10~keV~(a), 10--20~keV~(b), 20--40~keV~(c), 40--79~keV~(d).}
\label{fig:pp0406}
\end{figure*}

\noindent
In order to determine the spin period of LMC\,X-4 we have used the lightcurves in the quiescent state (observations 30102041004 and 30102041006). The true value of spin period and its uncertainty were obtained through inferring the sample population ($N=10000$) from original observational data using the Monte-Carlo algorithm \citep[see for the details][]{2013AstL...39..375B}. The resulting spin period value of the neutron star at the moment of \textsl{NuSTAR} observations \mbox{$P_{\rm spin} = 13.50124 \pm 0.00005 \text{ sec}$} is in good agreement with the values measured earlier with \textsl{NuSTAR} \citep{2017AstL...43..175S} and XRT/{\it Swift} \citep{2017MNRAS.464.2039M}. This may be the evidence of the absense of a significant acceleration/deceleration of the neutron star rotation in the LMC\,X-4 system. The value was used in the subsequent analysis.

\begin{figure*}

\begin{picture}(0,450)

\put(  0, 260){\includegraphics{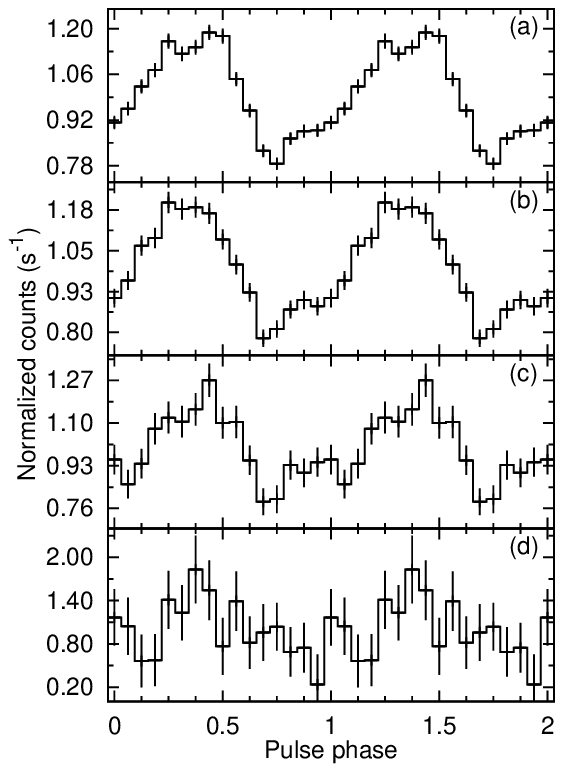}}
\put(165, 260){\includegraphics{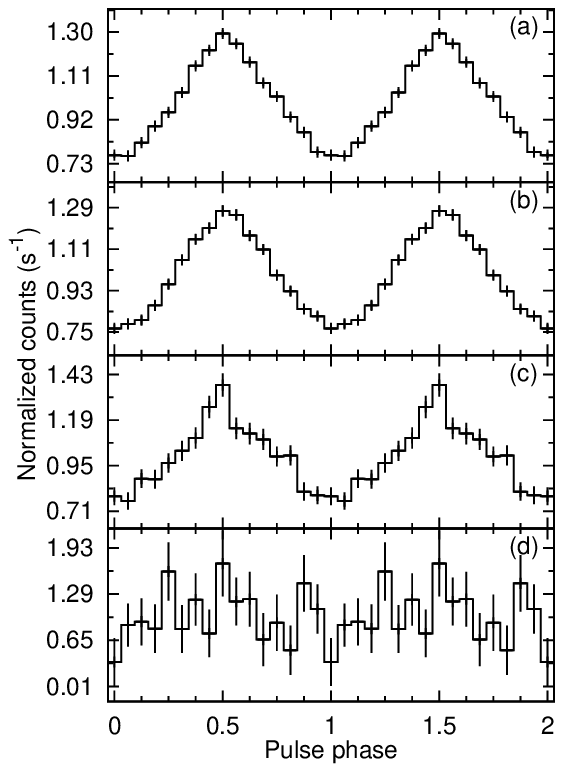}}
\put(330, 260){\includegraphics{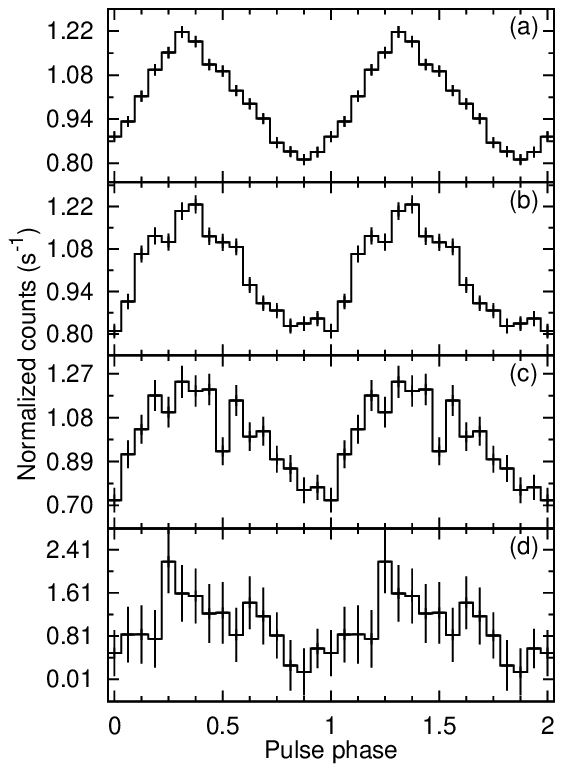}}

\put(  0,   0){\includegraphics{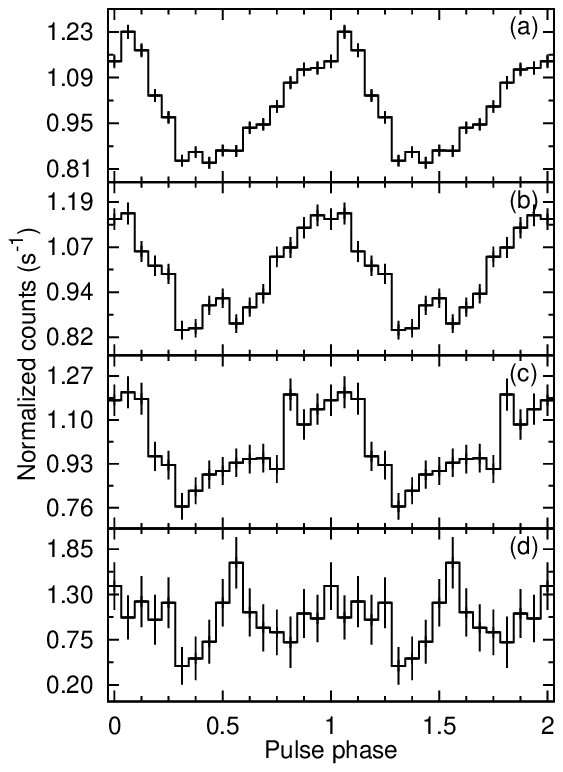}}
\put(165,   0){\includegraphics{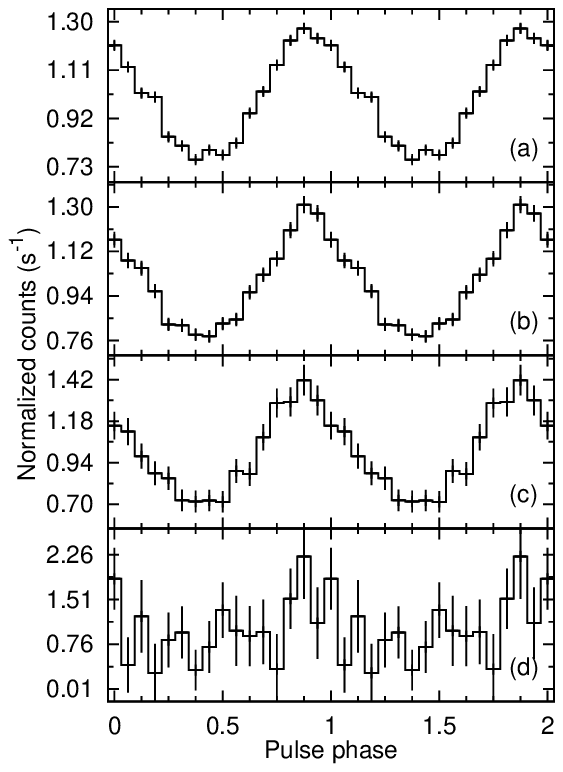}}
\put(330,   0){\includegraphics{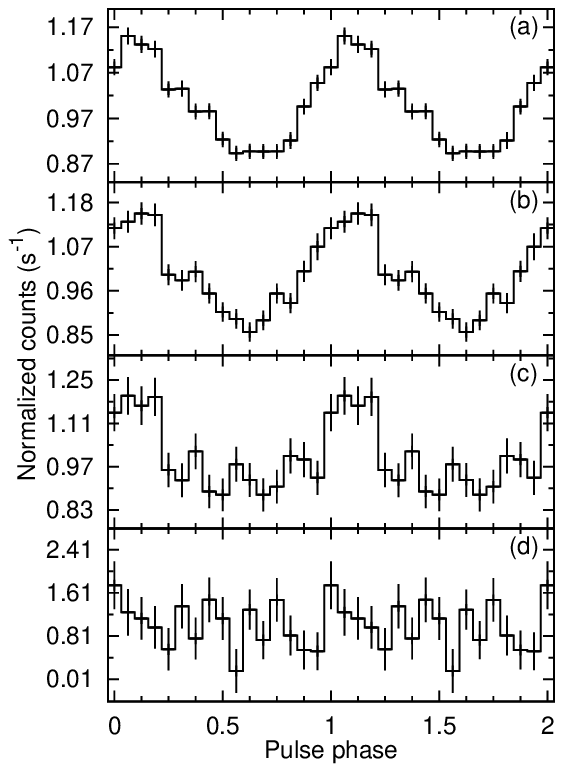}}

\put( 60, 495){interval (1)    }
\put(220, 495){interval (2-1) *}
\put(385, 495){interval (2-2)  }

\put( 55, 235){interval (3)    }
\put(220, 235){interval (4-1) *}
\put(385, 235){interval (4-2)  }
\end{picture}

\caption{Pulse profiles of LMC\,X-4 in the intervals (1), (2-1), (2-2), (3), (4-1) and (4-2) of the observation 30102041002 in the energy ranges: 3--10~keV~(a), 10--20~keV~(b), 20--40~keV~(c), 40--79~keV~(d). An intervals, containing flares, is labeled with an asterisk (*).}
\label{fig:pp021}

\end{figure*}

\begin{figure*}

\begin{picture}(0,450)

\put( 50, 260){\includegraphics{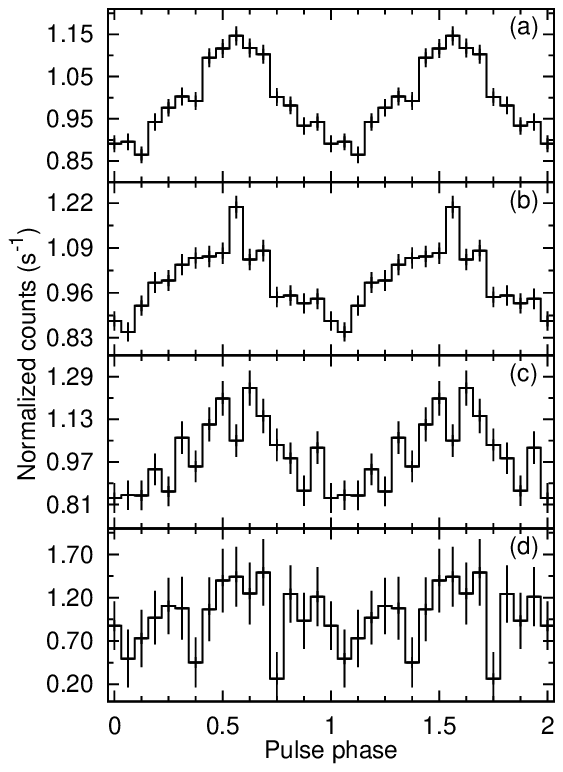}}
\put(260, 260){\includegraphics{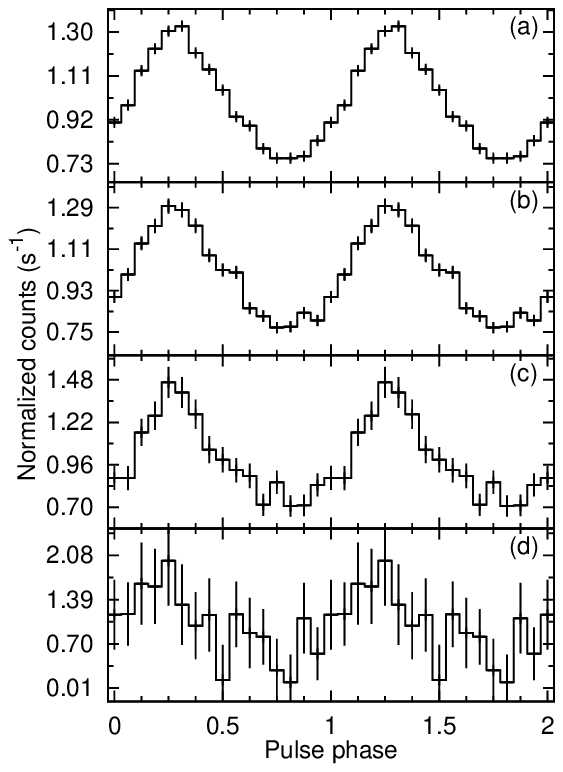}}

\put( 50,   0){\includegraphics{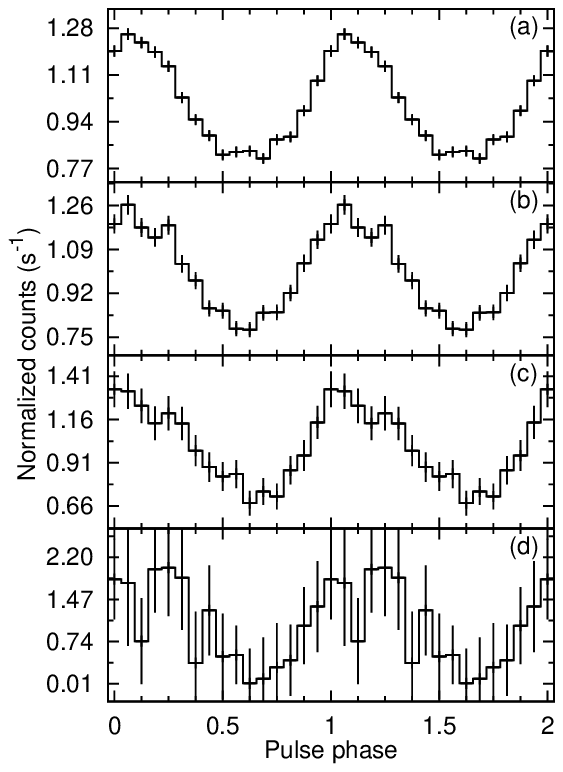}}
\put(260,   0){\includegraphics{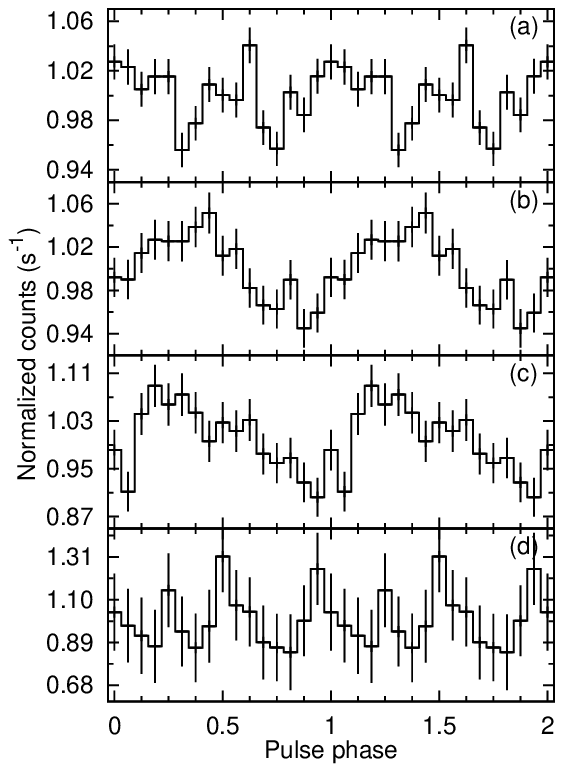}}

\put(110, 495){interval (5)    }
\put(315, 495){interval (6-1) *}

\put(105, 235){interval (6-2)  }
\put(315, 235){interval (7)    }
\end{picture}

\caption{Pulse profiles of LMC\,X-4 in the intervals (5), (6-1), (6-2) and (7) of the observation 30102041002 in the energy ranges: 3--10~keV~(a), 10--20~keV~(b), 20--40~keV~(c), 40--79~keV~(d). An intervals, containing flares, is labeled with an asterisk (*).}
\label{fig:pp022}

\end{figure*}

\begin{figure*}

\begin{picture}(0,450)

\put(  0, 260){\includegraphics{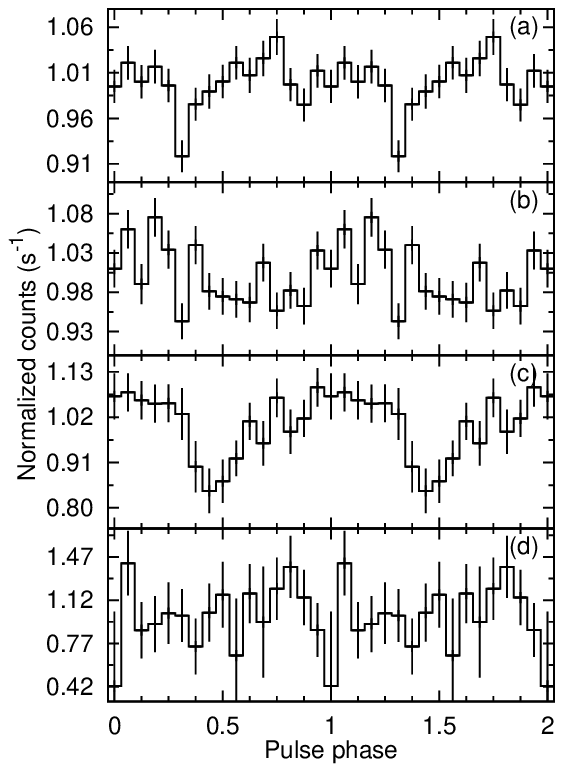}}
\put(165, 260){\includegraphics{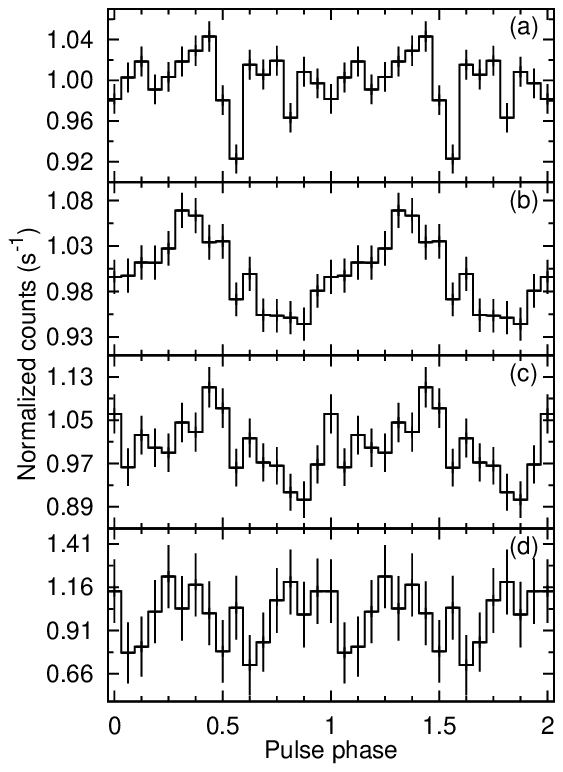}}
\put(330, 260){\includegraphics{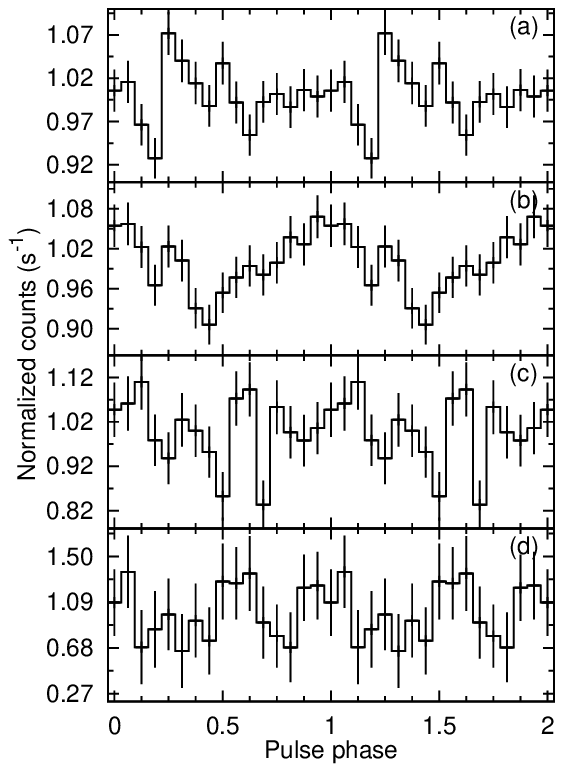}}

\put( 50,   0){\includegraphics{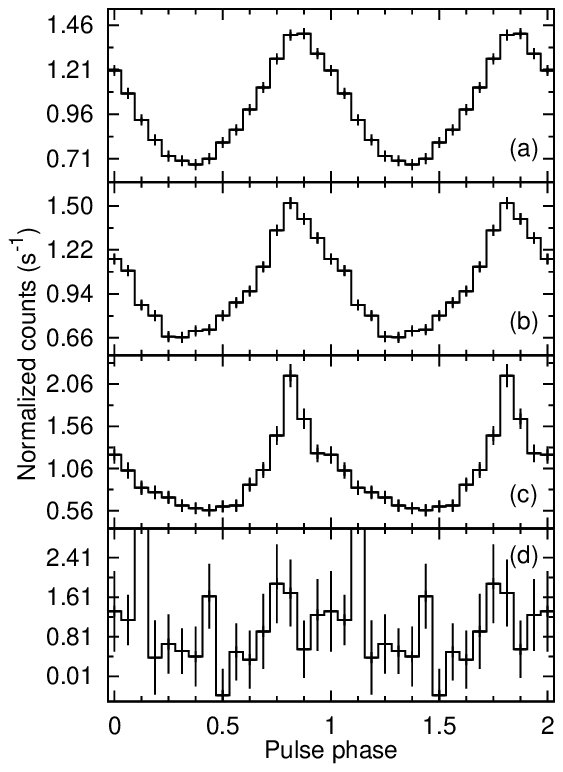}}
\put(260,   0){\includegraphics{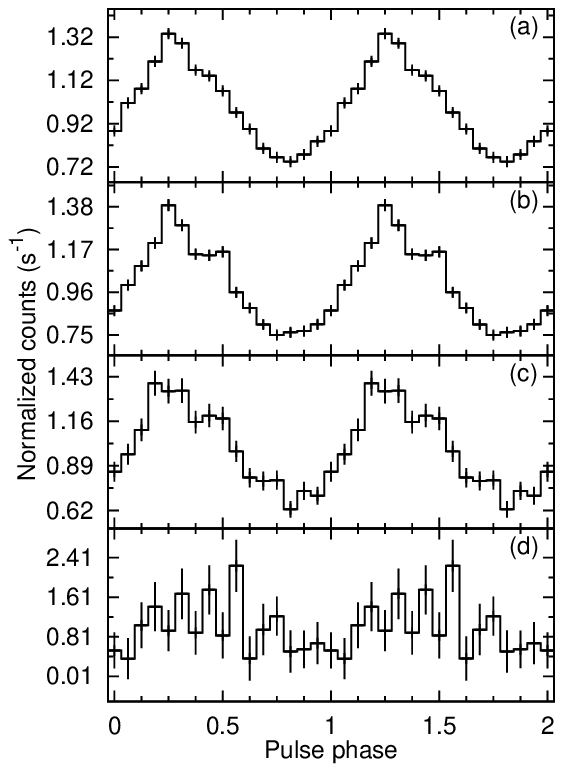}}

\put( 60, 495){interval (1)    }
\put(225, 495){interval (2)    }
\put(390, 495){interval (3)    }

\put(105, 235){interval (4-1) *}
\put(315, 235){interval (4-2)  }
\end{picture}

\caption{Pulse profiles of LMC\,X-4 in the intervals (1), (2), (3), (4-1) and (4-2) of the observation 30102041008 in the energy ranges: 3--10~keV~(a), 10--20~keV~(b), 20--40~keV~(c), 40--79~keV~(d). An intervals, containing flares, is labeled with an asterisk (*).}
\label{fig:pp08}

\end{figure*}

It is well-known that the pulse profile carries an important information about the geometry and physical properties of pulsar's emitting regions. A comparative analysis of the pulse profile and its evolution in flares and quiescent states allows one to explore the processes during the pulsar flaring activity. Figures \ref{fig:pp0406}, \ref{fig:pp021}, \ref{fig:pp022} and \ref{fig:pp08} show the evolution of pulse profiles in different energy ranges 3--10~keV, 10--20~keV, 20--40~keV and 40--79~keV for the chosen intervals. Note, that all the pulse profiles were built using the same value of zero epoch $T_{0} = 57325\text{ MJD}$.

\begin{figure*}
\vspace{0.13mm}

\centering

\begin{adjustbox}{max size={.67\textwidth}{.67\textheight}}
\begin{subfigure}{0.4\textwidth}
\includegraphics{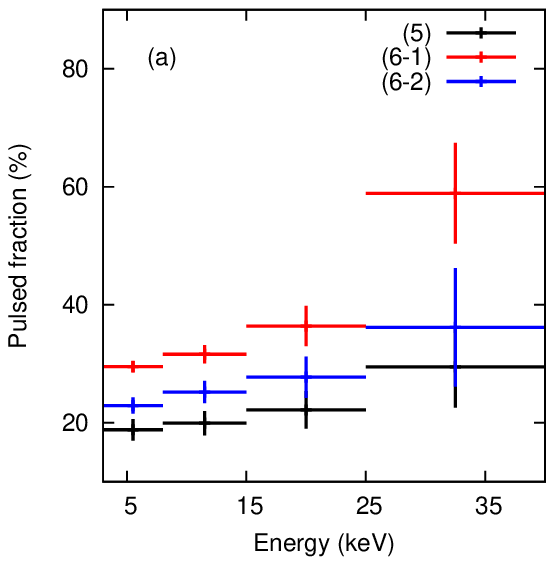}
\end{subfigure}
\begin{subfigure}{0.4\textwidth}
\includegraphics{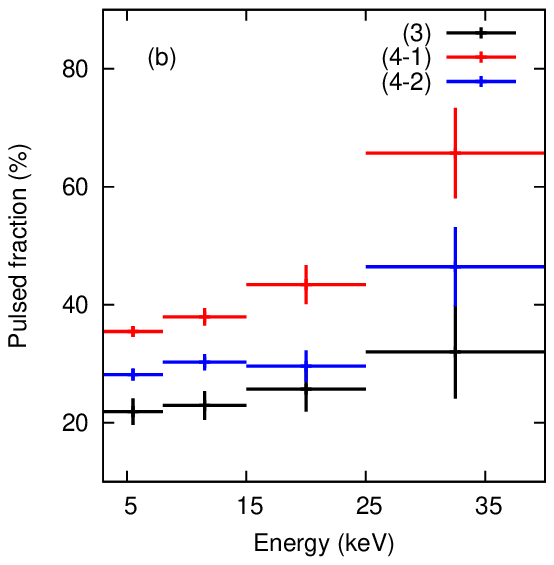}
\end{subfigure}
\end{adjustbox}

\begin{adjustbox}{max size={.67\textwidth}{.67\textheight}}
\begin{subfigure}{0.4\textwidth}
\includegraphics{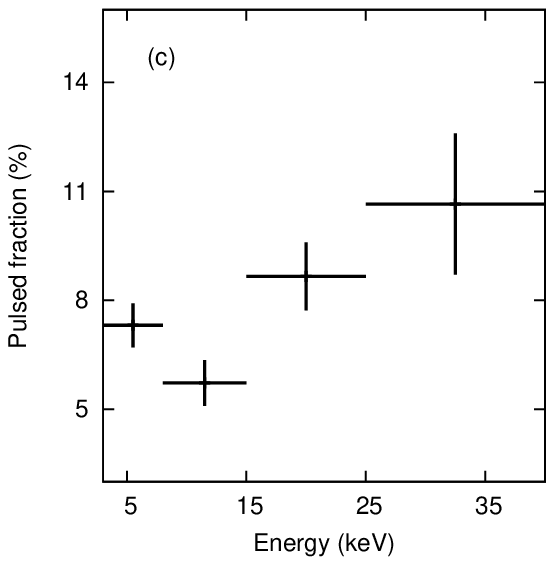}
\end{subfigure}
\begin{subfigure}{0.4\textwidth}
\includegraphics{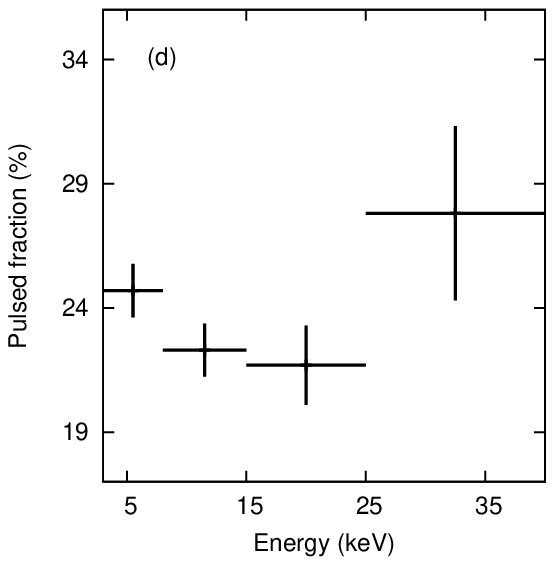}
\end{subfigure}
\end{adjustbox}

\caption{Pulsed fraction dependence on the energy in observations: in the intervals (5), (6-1) and (6-2) of the observation 30102041002 (panel a); in the intervals (3), (4-1) and (4-2) of the observation 30102041008 (panel b); in the observation 30102041004 (panel c); in the observation 30102041006 (panel d).}
\label{fig:pf}
\end{figure*}

In the absence of the flaring activity, in the vicinity of the maximum of the superorbital period (observation 30102041004), pulse profile has a complex shape consistent with previous results \citep[see e.g.][]{2000ApJ...541..194L,2017AstL...43..175S}. At the approach to the minimum of the superorbital period (observation 30102041006) the profile is simplified and assume a nearly sinusoidal shape (it is especially seen in the energy ranges of 3--10~keV and 10--20~keV). Simultaneously pulsed fraction increases by a factor of~$\sim 3$ (Fig.~\ref{fig:pf}).

Pulse profile during flares reach a nearly triangular shape, having inbetween a simple assymetric shape, similar to the sinusoidal one without pronounced features. After flaring activity episode pulse profile returns back to a more complex shape, similar to one seen during the maximum of the superorbital period (see e.g. interval 7 in the observation 30102041002 and intervals 1, 2, 3 in the observation 30102041008). Another important fact is the significant drift of the pulse phase (up to 100\% in its maximum).



The pulsed fraction is also significantly varied with the energy during the flares for different observational intervals (Fig.~\ref{fig:pf}). One can see from this figure that the pulsed fraction raises during the flares up to $\sim 60-70\%$ in the energy range 25--40\,keV, which is much higher than the quiescent state values $\sim 6-14\%$ \citep{2005AstL...31..380T,2017AstL...43..175S}. In the afterglow and just before the flare pulsed fraction also reaches higher values: $25-50\%$ and $20-30\%$, correspondingly. It is worth noting that these changes are observed for all the flares analyzed in the paper. Also it is necessary to note that in the observation 30102041006, when the system was in the low state of the superorbital cycle, the pulsed fraction is also sufficiently higher than in the high state and reaches values $\sim 20-30\%$ even in the absence of flares. It was shown by \citet{2009AstL...35..433L} that the pulsed fraction for bright X-ray pulsars is usually decreased when the luminosity increases. However, when it approaches the Eddington limit, the pulsed fraction can increase again with the luminosity increase, which apparently is seen in the LMC\,X-4 system.

\bigskip

\subsection*{Energy spectra}
\noindent
As in our previous paper \citep{2017AstL...43..175S} a thermal comptonization model {\sc comptt} \citep{1994ApJ...434..570T} was used to approximate spectra of \mbox{LMC\,X-4}. A crucial advantage of this model is that it fits well the spectrum of LMC\,X-4 and has a physical justification. The following components has been added to improve the quality of the fit: {\sc phabs} to account for an interstellar absorption and {\sc gaus} to account for a fluorescent $K_{\alpha}$ iron line at 6.4\,keV. Spectra from FPMA and FPMB modules were analyzed simultaneously, a normalization coefficient $C$ was introduced to account the different calibrations of the modules. All other parameters were fixed between the data sets. A fit quality was estimated using $\chi^2$ per degree of freedom (d.o.f.). The hydrogen column density was fixed at the value of $N_{\rm H} = 5.74 \times 10^{20} \text{ atoms cm}^{-2}$ \citep[assuming a solar abundance,][]{1990ARA&A..28..215D}. The best-fit parameters are presented in Table~\ref{tab:fits}. It can be seen that the proposed model describes the pulsar spectra well in all observed states.

\begin{figure*}
\vspace{15mm}

\begin{adjustbox}{max size={.8\textwidth}{.8\textheight}}
\begin{picture}(0,450)
\put(0  , 270){\includegraphics{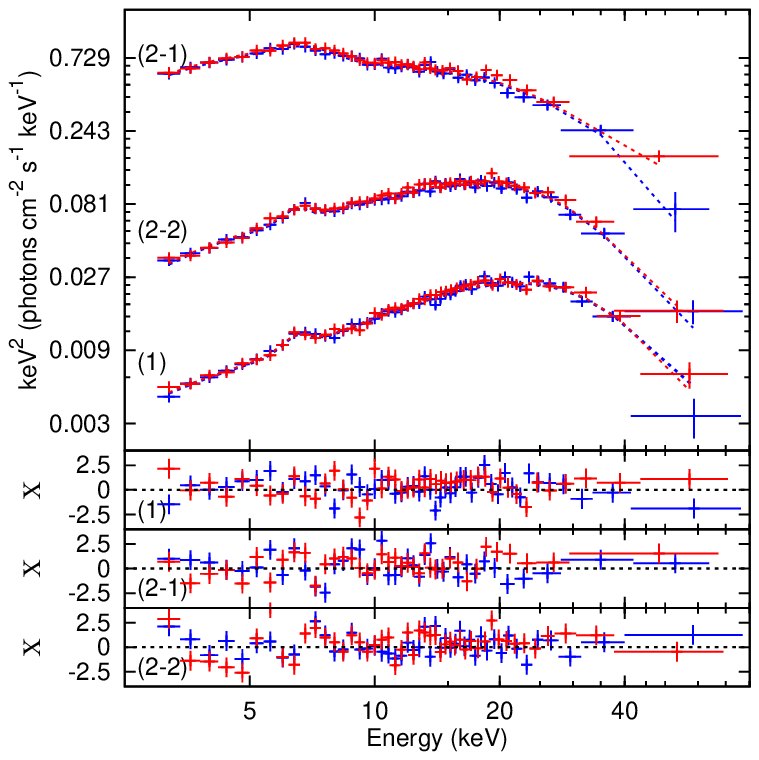}}
\put(250, 270){\includegraphics{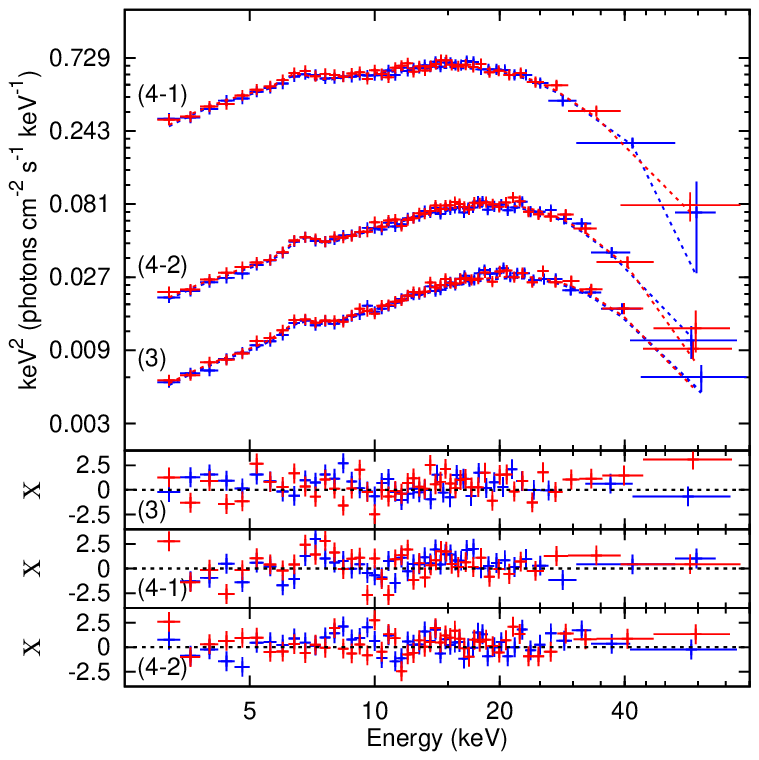}}

\put(0  ,  20){\includegraphics{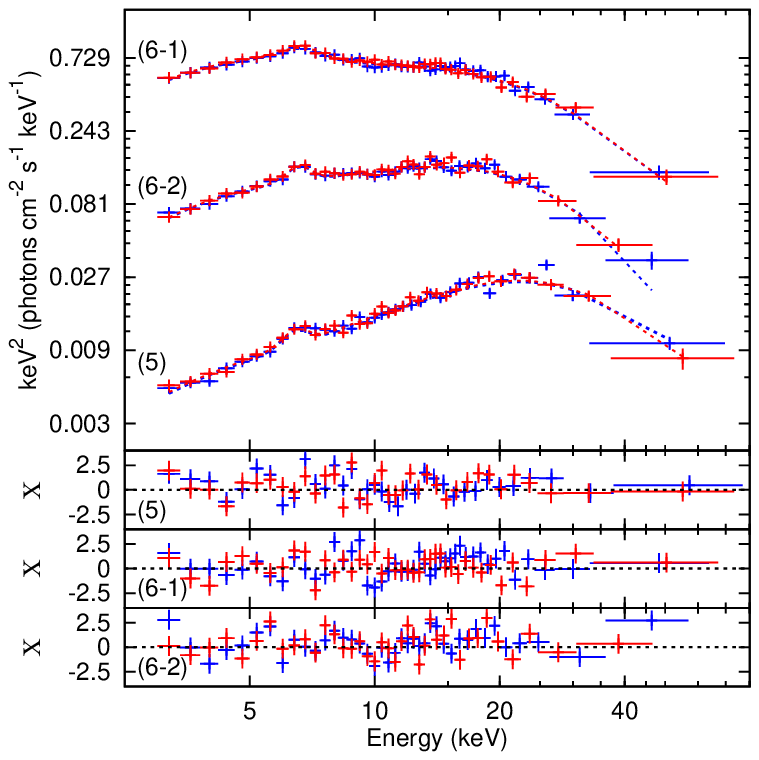}}
\put(250,  20){\includegraphics{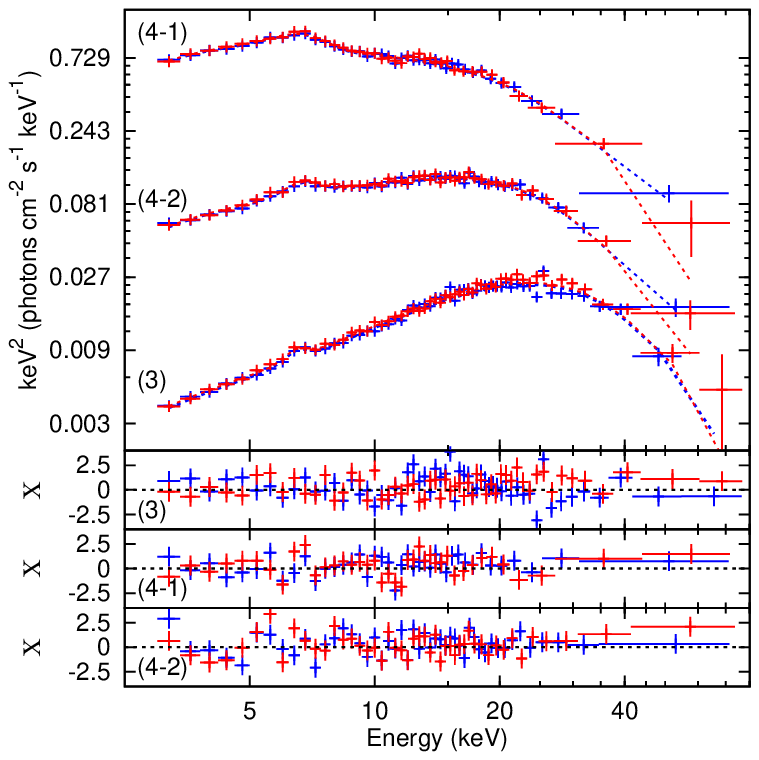}}

\put(44 , 495){flare 1 (FL1, 30102041002)}
\put(294, 495){flare 2 (FL2, 30102041002)}

\put(44 , 245){flare 3 (FL3, 30102041002)}
\put(294, 245){flare 4 (FL4, 30102041008)}
\end{picture}
\end{adjustbox}

\caption{Energy spectra of LMC\,X-4 before, during and after the flares. Spectra in intervals before the flare plotted using coefficient 0.1, after the flare -- using coefficient 0.2. Labels of intervals follow the numeration presented in Fig.~\ref{fig:lc02} and Fig.~\ref{fig:lc08}. Corresponding residuals are shown in the bottom panels.}
\label{fig:spectra}

\end{figure*}

\begin{figure*}
\centering

\includegraphics{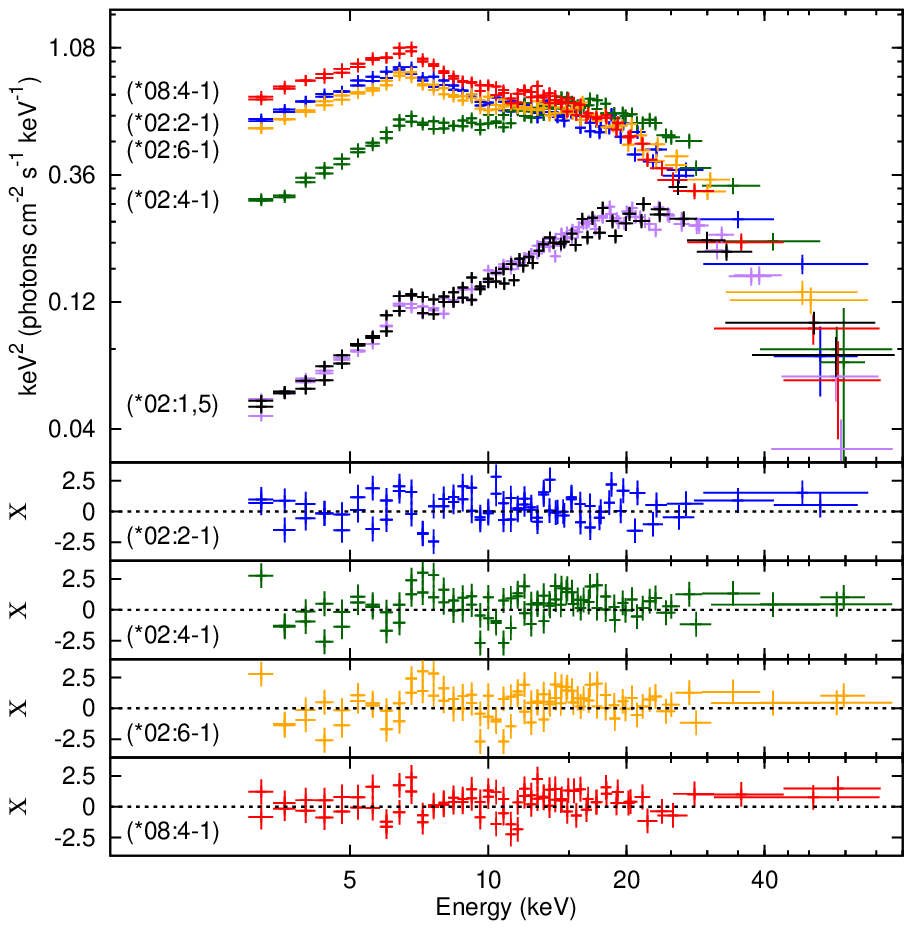}

\caption{Superimposed spectra of flares in the intervals (2-1), (4-1) and (6-1) of the observation 30102041002 (*02:2-1, *02:4-1 and *02:6-1), in the interval (4-1) of the observation 30102041008 (*08:4-1) and spectra in quiescent states in the intervals (1) and (5) of the observation 30102041002 (*02:1,5). Corresponding residuals are shown in the bottom panels.}
\label{fig:spectra_allflare}
\end{figure*}

Figure~\ref{fig:spectra} shows the evolution of \mbox{LMC\,X-4} spectra before, during and after the flares, plotted together for each of the four flares. For illustrative purposes the spectra plotted with the different coefficients: 0.1 before the flares and 0.2 after. The figure shows that during the flares spectra softens significantly. The emission softening during flares in \mbox{LMC\,X-4} was noted previously in a number of papers \citep[see e.g.][]{1989ESASP.296...39D,1991ApJ...381..101L,2000ApJ...541..194L,2003ApJ...586.1280M}. Overall estimate shows that the most significant change in the spectra appear at low energies. To verify this assumption, we plotted together spectra during flares and in quiescense (Fig.~\ref{fig:spectra_allflare}). It can be seen clearly that the spectra differ mainly up to $\sim 20 \text{ keV}$, while at higher energies they remain practically identical.

Figure~\ref{fig:pop_0208} shows the evolution of the spectral parameters in the intervals (1)--(7) of the observation 30102041002 and in the intervals (1)--(4) of the observation 30102041008. There is a significant decrease (by~$\sim 1.5-2$~times) in the plasma optical depth of the emitting region during flares, while the plasma temperature is decreased during afterglows. One also can see the increment of the seed soft photon temperature during the flare series, that ceases after it ends. It is worth noting that at least during the first flare in series (FL1, FL4) the equivalent width of the fluorescent iron line $EW_{\rm Fe}$ is increased by a factor of~$\sim 3$.


\section*{Discussion and Conclusion}

\begin{table*}
\centering
\caption{ Parameters of the best fit models for the spectra of LMC\,X-4, obtained in different states.}
\setlength\tabcolsep{3pt}
\begin{tabular}{c|c|c|c|c|c|c|c|c|c|c}
\hline\hline
Obs.      & Int. & $C$ & $T_0, $     & $kT$,          & $\tau$            & $E_{\rm Fe}$,     & $\sigma_{\rm Fe}$, & $EW_{\rm Fe}$,   & $f_{\rm x} \times 10^{-9}$, & $\chi^2/d.o.f$ \\
          &      &     & keV         & keV            &                   & keV               & keV                & eV               &  ${\rm erg/cm^2/s}$      & \\
\hline
\multirow{10}{*}{*02}
 & (1)     & 1.02 & $0.93 \pm 0.09$  & $8.78 \pm 0.10$ & $12.29 \pm 0.17$ & $6.45 \pm 0.05$ & $0.44 \pm 0.09$ & $254 \pm  28$ & 0.68                 & 0.995          \\
 & (2-1)   & 1.03 & $1.08 \pm 0.06$  & $8.78 \pm 0.40$ & $ 6.32 \pm 0.32$ & $6.32 \pm 0.14$ & $1.19 \pm 0.16$ & $644 \pm 146$ & 2.47                 & 1.001          \\
 & (2-2)   & 1.02 & $1.00 \pm 0.06$  & $7.60 \pm 0.10$ & $11.24 \pm 0.16$ & $6.54^*       $ & $0.60^* 	    $ & $340 \pm  25$ & 1.60                 & 1.016          \\
 & (3)     & 1.02 & $1.15 \pm 0.05$  & $8.93 \pm 0.11$ & $11.40 \pm 0.19$ & $6.51 \pm 0.05$ & $0.53 \pm 0.10$ & $265 \pm  31$ & 0.76                 & 1.081          \\
 & (4-1)   & 1.02 & $1.16 \pm 0.04$  & $7.64 \pm 0.13$ & $ 9.45 \pm 0.18$ & $6.61^*       $ & $0.64^*  	    $ & $307 \pm  29$ & 2.03                 & 1.127          \\
 & (4-2)   & 1.03 & $1.05 \pm 0.07$  & $8.00 \pm 0.10$ & $11.66 \pm 0.19$ & $6.63 \pm 0.04$ & $0.43 \pm 0.07$ & $252 \pm  26$ & 1.12                 & 1.024          \\
 & (5)     & 1.02 & $1.02 \pm 0.08$  & $9.04 \pm 0.12$ & $11.89 \pm 0.19$ & $6.50^*       $ & $0.51^*       $ & $336 \pm  31$ & 0.70                 & 0.951          \\
 & (6-1)   & 1.01 & $1.12 \pm 0.05$  & $8.68 \pm 0.29$ & $ 6.54 \pm 0.24$ & $6.47 \pm 0.08$ & $0.79 \pm 0.14$ & $385 \pm  55$ & 2.43                 & 0.920          \\
 & (6-2)   & 1.01 & $1.18 \pm 0.08$  & $7.39 \pm 0.19$ & $ 9.18 \pm 0.35$ & $6.52 \pm 0.08$ & $0.42 \pm 0.18$ & $276 \pm  41$ & 2.27                 & 1.161          \\
 & (7)     & 1.03 & $0.86 \pm 0.06$  & $9.29 \pm 0.05$ & $13.43 \pm 0.89$ & $6.46 \pm 0.04$ & $0.44 \pm 0.05$ & $215 \pm  19$ & 0.98                 & 1.003          \\
\hline
*04 & -     & 1.03 & $1.04 \pm 0.05$ & $8.81 \pm 0.05$ & $14.46 \pm 0.14$ & $6.53 \pm 0.04$ & $0.43 \pm 0.07$ & $180 \pm 13$ & 1.17                 & 1.060          \\
\hline
*06 & -     & 1.03 & $0.86 \pm 0.16$ & $8.30 \pm 0.08$ & $16.53 \pm 0.27$ & $6.28 \pm 0.06$ & $0.53 \pm 0.08$ & $365 \pm 29$ & 0.35                 & 0.991          \\
\hline
\multirow{4}{*}{*08}
& (1)   & 1.04 & $0.68 \pm 0.13$ & $9.31 \pm 0.08$ & $13.51 \pm 0.14$ & $6.28 \pm 0.09$ & $0.67 \pm 0.11$ & $291 \pm 38$ & 0.70 & 0.959 \\ & (2)   & 1.03 & $0.87 \pm 0.08$ & $9.07 \pm 0.08$ & $13.81 \pm 0.14$ & $6.39 \pm 0.05$ & $0.28 \pm 0.08$ & $145 \pm 17$ & 0.69 & 1.071 \\ & (3)   & 1.03 & $0.90 \pm 0.08$ & $9.11 \pm 0.09$ & $12.96 \pm 0.16$ & $6.46 \pm 0.08$ & $0.36 \pm 0.12$ & $154 \pm 24$ & 0.64 & 1.044 \\ & (4-1) & 1.01 & $1.05 \pm 0.05$ & $7.52 \pm 0.24$ & $ 6.81 \pm 0.25$ & $6.32 \pm 0.10$ & $0.96 \pm 0.11$ & $568 \pm 78$ & 2.72 & 0.971 \\ & (4-2) & 1.02 & $1.00 \pm 0.04$ & $7.29 \pm 0.09$ & $ 9.70 \pm 0.13$ & $6.61^*       $ & $0.67^*       $ & $381 \pm 27$ & 1.90 & 0.969 \\

\hline
\multicolumn{11}{l}{}\\
\multicolumn{11}{l}{$^*$ \text{ -- parameter poorly contrained and was fixed}} \\

\end{tabular}

\label{tab:fits}
\end{table*}

\noindent
Episodes of flaring activity of LMC\,X-4 are known for a long time. It is a series of short-term increases in the source luminosity occuring approximately once a day \citep{1991ApJ...381..101L,2000ApJ...541..194L,2003ApJ...586.1280M}, however the regularity and flaring mechanism are still unclear. Several models of flares in LMC\,X-4 has been proposed, nevertheless having its own drawbacks each.

\bigskip

\subsection*{Flaring activity}

\begin{figure*}
\centering

\begin{adjustbox}{max size={.66\textwidth}{.66\textheight}}
\begin{subfigure}{0.4\textwidth}
\includegraphics{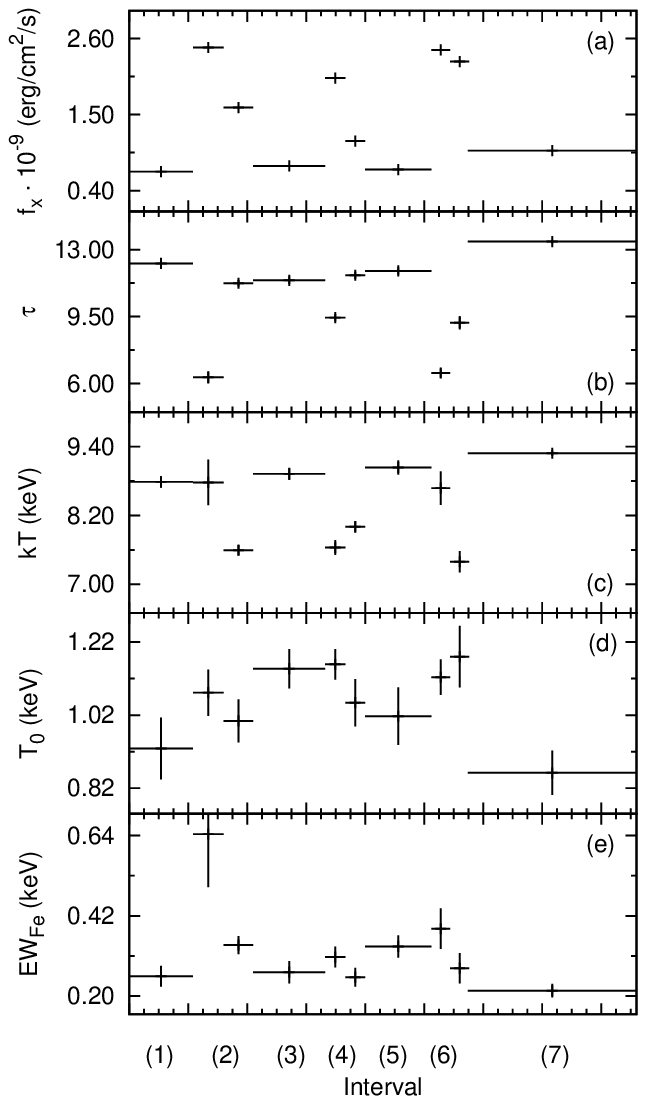}
\end{subfigure}
\begin{subfigure}{0.4\textwidth}
\includegraphics{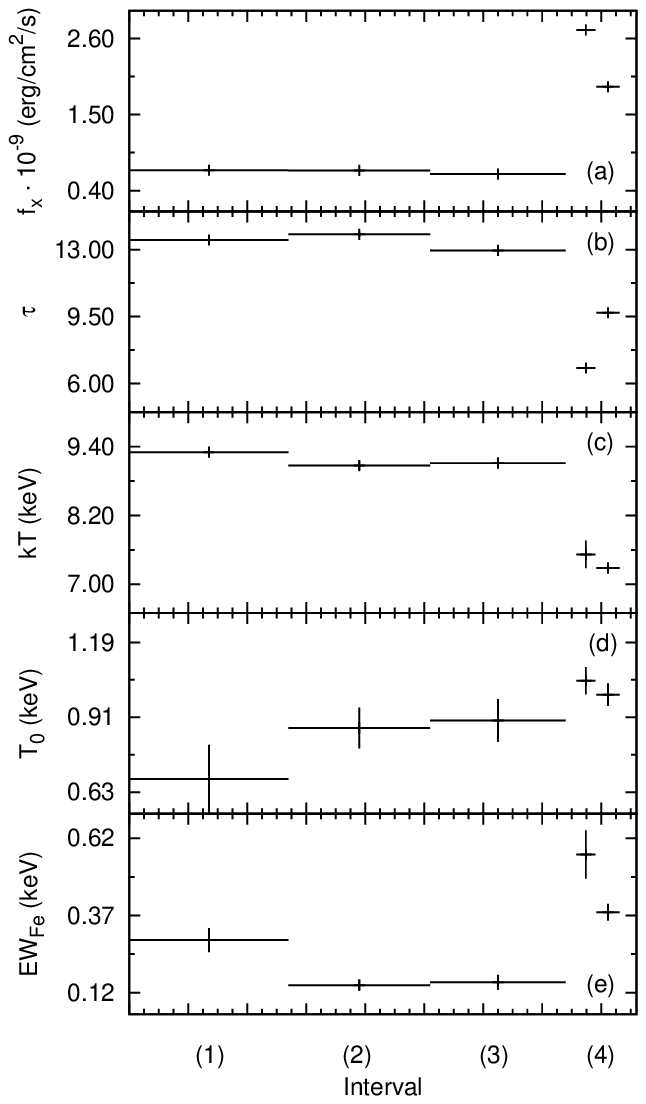}
\end{subfigure}
\end{adjustbox}

\caption{Evolution of spectral parameters during 30102041002 (left panel) and 30102041008 (right panel) observations: flux (a), optical depth (b) plasma temperature (c), seed photons temperature (d), equivalent width of the fluorescent iron line (e). Labels of intervals follow the numeration presented in Fig.~\ref{fig:lc02} and Fig.~\ref{fig:lc08}.}
\label{fig:pop_0208}
\end{figure*}

\noindent
As it has been noted in the Introduction, flares in systems with accreting neutron stars can be caused by different physical mechanisms, among which the following could be mentioned: thermonuclear burning on a surface of a neutron star (type I bursts); accretion flow variations (type II bursts); non-stationary accretion (similar to what observed in the so-called supergiant fast X-ray transients, SFXT); magnetospheric activity (magnetosphere perturbations) of a pulsar.

\citet{1998ApJ...496..915B} has shown that it is difficult to explain the observed flares of LMC\,X-4 in the framework of thermonuclear burning on the surface of a neutron star. Besides that, thermonuclear flashes are characterized by a spectral softening during the course of flare evolution, while in the case of LMC\,X-4 one can see the contrary behavior -- the spectral hardening during the flares evolution.

Variability of an accretion flow is another frequently considered mechanism for the flaring activity. Let us consider the simplest scenario in which the accretion flow variations supply an additional volume of matter, whose energy is released during the accretion and leads to the increased luminosity of the source. Assuming the efficiency of converting of the kinetic energy of the infalling matter to the emission is $\eta \sim 10 \%$ we estimate the amount of matter necessary to obtain the luminosities observed in flares FL3 (the average luminosity is $L_{x} =  7.27\times10^{38}$\,\ergs) and FL4 (the average luminosity is $L_{x} = 8.14\times10^{38}$\,\ergs). To accomplish it, we split the flares into 200~second intervals and calculated the flux in each of them using the spectral model discussed above. The resulting amount of matter was obtained by integrating over all time intervals. It is approximately $M \simeq 1.8\times10^{22} \text{ g}$ for the FL3 flare and $M \simeq 2.3\times10^{22}\text{ g}$ for the FL4 one. Taking into account the accretion rate in a quiescent state $\dot{m} \simeq 3.9\times10^{18} \text{ g s}^{-1}$, it is possible within this framework to explain the occurence of the LMC\,X-4 flares. However, a detailed study is required for a more reliable interpretation.

\citet{2000ApJ...541..194L} noted a certain similarity in the observed flaring activity of LMC\,X-4 (in particular, a super-Eddington character of flares) with the flares, observed in magnetars \citep{1995MNRAS.275..255T,2017ARA&A..55..261K}. However, a detailed examination reveals significant differences in the flaring activity of them. Magnetars are characterized by ultrashort flares (duration of $\sim 0.1-10 \text{ s}$), spectral hardening during the flares, absense of pulsations in flares peaks, large interflare periods, etc. This allows one to conclude that the magnetar mechanism is unsuitable for describing LMC\,X-4 flares \citep{2000ApJ...541..194L}, despite the fact that the magnetic field strength in the system can be sufficiently large \citep[$> 10^{13}\text{ G}$,][]{2017AstL...43..175S}.

Another mechanism, suitable for explaining the flaring activity in the LMC\,X-4 system, can be the modification of the model, proposed by \citet{2012MNRAS.420..216S,2013PhyU...56..321S} for explaining the phenomenon of SFXTs flares. Indeed, comparing the flaring activities in LMC\,X-4 and SFXTs, one may see a number of similarities -- flares shapes, characteristic time scales of flares evolution, etc. However, the flaring mechanism for SFXTs assumes quiescent luminosities lower than $L_x \sim 10^{36}$\,\ergs. Thus, it is hard to explain frequent flares with this model at higher luminosities, which is typical for LMC\,X-4.

\bigskip

\subsection*{Changes of the radiation properties with the luminosity increased to super-Eddington values}

\noindent
Due to a high sensivity and wide energy range of the \textsl{NuSTAR} obervatory it became possible for the first time to trace in great details changes in the spectrum and pulse profiles of LMC\,X-4 with the luminosity increased above the super-Eddington level. The source spectra both during flares and quiescent, are well approximated with the model of thermal comptonization ({\sc comptt}). During flares, with luminosities above $L_{\rm x} \sim 6 \times 10^{38}$\,\ergs, the following changes in the spectra and pulse profiles were observed:

\begin{enumerate}
\item pulse profile in the energy range 3--40~keV becomes approximately triangular (Fig.~\ref{fig:pp021},~\ref{fig:pp022},~\ref{fig:pp08});
\item pulsed fraction increases with the energy increase, reaching $60-70\%$ in the energy range 25--40~keV, which is much higher than the quiesent state values $6-14\%$ (Fig.~\ref{fig:pf});
\item with the luminosity increased by more than a factor of 10 spectra shapes changes mainly up to $\sim 20-30\text{ keV}$, while at higher energies both the shapes and the registered flux remain practically unchanged (see Fig.~\ref{fig:spectra},~\ref{fig:spectra_allflare});
\item a negative correlation between the plasma temperature and optical depth and the X-ray luminosity of the source (Table~\ref{tab:fits}, Fig.~\ref{fig:pop_0208}).
\end{enumerate}

It is well-known that many X-ray pulsars demostrate spectral hardening during flares. At the same time, above some critical luminosity (several $\times 10^{37}$\,\ergs), a saturation and even a spectral softening of a pulsar spectrum may occur \citep{2015MNRAS.452.1601P}. These authors proposed a model of an accretion column where such changes are explained by a Compton saturation, at which the comptonized emission from the wall of an accretion column is reflected in the atmosphere of the neutron star \citep{2013ApJ...777..115P}. With the increase of the accretion rate, the height of the column increases and the fraction of a reflected, predominantly hard, spectral component begin to decrease, which leads to a pulsar spectrum softening. Moreover, according to the standard model of the accretion column \citep{1975A&A....42..311B,1976MNRAS.175..395B}, the column height increases with increasing luminosity and a fan-beam radiation pattern becomes dominant. For a distant observer it leads to the fact that the pulse profile becomes double-peaked \citep{1973A&A....25..233G}. It is notable that qualitatively similar behaviour can also be explained in the framework of the modern model for the formation of an X-ray pulsar radiation, which takes into account the reflection of the accretion column radiation from the surface of the neutron star (see \citealt{2013ApJ...777..115P} and Fig.~6 in the paper of \citealt{2015MNRAS.448.2175L}).

LMC\,X-4 exhibit a spectral softening with the luminosity increase, similar to that mentioned in the paper of \citet{2015MNRAS.452.1601P}. However, the registered pulse profile -- single-peaked, triangular -- is more likely indicate a pencil-beam radiation pattern, which is typical for a low height accretion column. In this case one can assume the distortion/disappearance of the column at the super-Eddington luminosities. At the same time, the observed decrease of the plasma temperature and optical depth are also consistent with an accretion column height decrease, bearing in mind that, in the framework of \citet{2015MNRAS.452.1601P} model, the comptonizing medium is the electrons in the outermost edges of the accretion column. Another possible explanation for the observed change in the pulse profiles is the transition of LMC\,X-4 into the state of ultraluminous X-ray source (ULX), which are also characterized by single-peaked and approximately sinusoidal pulse profiles along with the pulsed fraction increase with energy \citep{2016ApJ...831L..14F,2017MNRAS.467.1202M}.

\section*{Acknowledgements}

\noindent
AS, AL and SM would like to thank the grant RNF~14-12-01287. In this paper the \textsl{NuSTAR} observatory, a project of University of California, created with the support of NASA and NASA/JPL, data was used. Data was processed with the {\sc nustardas} software, which is joint development product of ASDC (Italy) and Caltech (USA).

\nocite{*} 
\bibliographystyle{mnras}

\bibliography{lmcx4_nustar_flares} 

\end{document}